\documentclass{PoS}

  \usepackage{amssymb}
   \usepackage{amsmath}
    \usepackage{amsfonts}
     \usepackage{epsfig}
      \usepackage{bm}

\usepackage{mathpazo}

\title{Diffractive production of heavy mesons at the LHC within $k_t$ - factorization approach}

\ShortTitle{Diffractive production of heavy mesons}

\author{\speaker{Marta Luszczak}%
        \thanks{This work was partially supported by the Polish National Science Centre grant DEC-2013/09/D/ST2/03724.}\\
       University of Rzesz\'ow, PL-35-959 Rzesz\'ow, Poland\\
       E-mail: \email{luszczak@ur.edu.pl}}

\author{Antoni Szczurek\\%
        Institute of Nuclear Physics PAN, PL-31-342 Cracow, Poland\\
       E-mail: \email{antoni.szczurek@ifj.edu.pl}}

\abstract{We discuss diffractive production of heavy mesons at the LHC 
\cite{Luszczak:2014cxa, Luszczak:2016csq}.
The differential cross sections for single- and central-diffractive 
mechanisms for $c\bar c$ pair production are calculated in the framework
of the Ingelman-Schlein model corrected for absorption effects. 
Here, leading-order gluon-gluon fusion and quark-antiquark anihilation
partonic subprocesses are taken into consideration. 
Both pomeron flux factors as well as parton distributions in the pomeron are
taken from the H1 Collaboration analysis of diffractive structure function and diffractive
dijets at HERA. The extra corrections from subleading reggeon exchanges
are also taken into consideration. 
In addition to standard collinear approach, for the first time 
the differential cross sections for the diffractive $c\bar c$
pair production are calculated in the framework of the
$k_t$-factorization approach, i.e. effectively including higher-order
corrections. The unintegrated (transverse momentum dependent) diffractive
parton distributions in proton are calculated with the help of the Kimber-Martin-Ryskin
prescription where collinear diffractive PDFs are used as input.
Some correlation observables, like azimuthal angle correlation between 
$c$ and $\bar c$, and $c \bar c$ pair transverse momentum were obtained
for the first time.  
The hadronization of charm quarks is taken into account by means of fragmentation function
technique.}

\FullConference{XXIV International Workshop on Deep-Inelastic Scattering and Related Subjects\\
		 11-15 April, 2016\\
		 DESY Hamburg, Germany}

\begin{document}

\section{Introduction}

Diffractive hadronic processes  were studied theoretically in the
so-called resolved pomeron model \cite{IS}. 
During the studies performed at Tevatron, it was realized that the
model, previously used to describe deep-inelastic diffractive processes 
must be corrected to take into account absorption effects related to
hadron-hadron interactions. 
Such interactions, unavoidably present in hadronic collisions at high
energies are not present in electron/positron induced processes. 
In theoretical models this effect is taken into account by multiplying 
the diffractive cross section calculated using HERA diffractive PDFs 
by a kinematics independent factor called the gap survival probability -- $S_{G}$. 
Two theoretical groups specialize in calculating such probabilities
\cite{KMR2000, Maor2009}. 
At high energies such factors, interpreted as probabilities, are 
very small (of the order of few \%). This causes that the predictions 
of the diffractive cross sections are not as precise as those for 
the standard inclusive non-diffractive cases. This may become a
challenge when a precise data from the LHC will become available. 

In this study we consider diffractive production of charm for which
rather large cross section at the LHC are expected, 
even within the leading-order (LO) collinear approach
\cite{Luszczak:2014cxa}. 
On the other hand, it was shown that for the inclusive non-diffractive 
charm production the LO collinear approach is a rather poor
approximation and higher-order corrections are crucial. 
Contrary, the $k_t$-factorization approach, which effectively includes 
higher-order effects, gives a good description of the LHC data  
for inclusive charm production at $\sqrt{s}$ = 7 TeV 
(see \textit{e.g}. Ref.~\cite{Maciula:2013wg}). 
This strongly suggests that application of $k_t$-factorization approach 
to diffractive charm production is useful. 
This presentation is based on our recent study in \cite{Luszczak:2016csq}.

\section{Formalism}

\begin{figure}[!htbp]
\begin{minipage}{0.4\textwidth}
 \centerline{\includegraphics[width=1.0\textwidth]{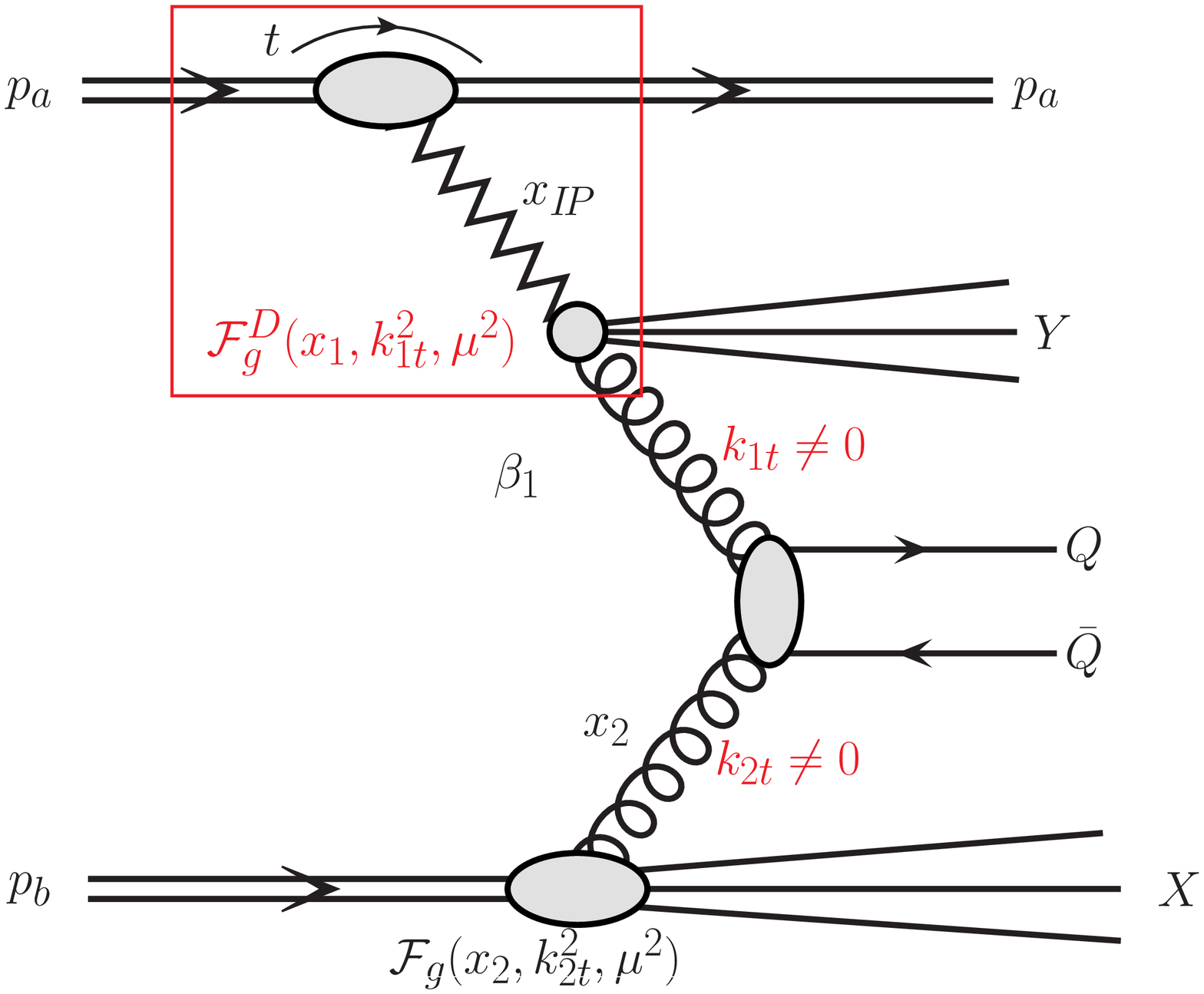}}
\end{minipage}
\begin{minipage}{0.4\textwidth}
 \centerline{\includegraphics[width=1.0\textwidth]{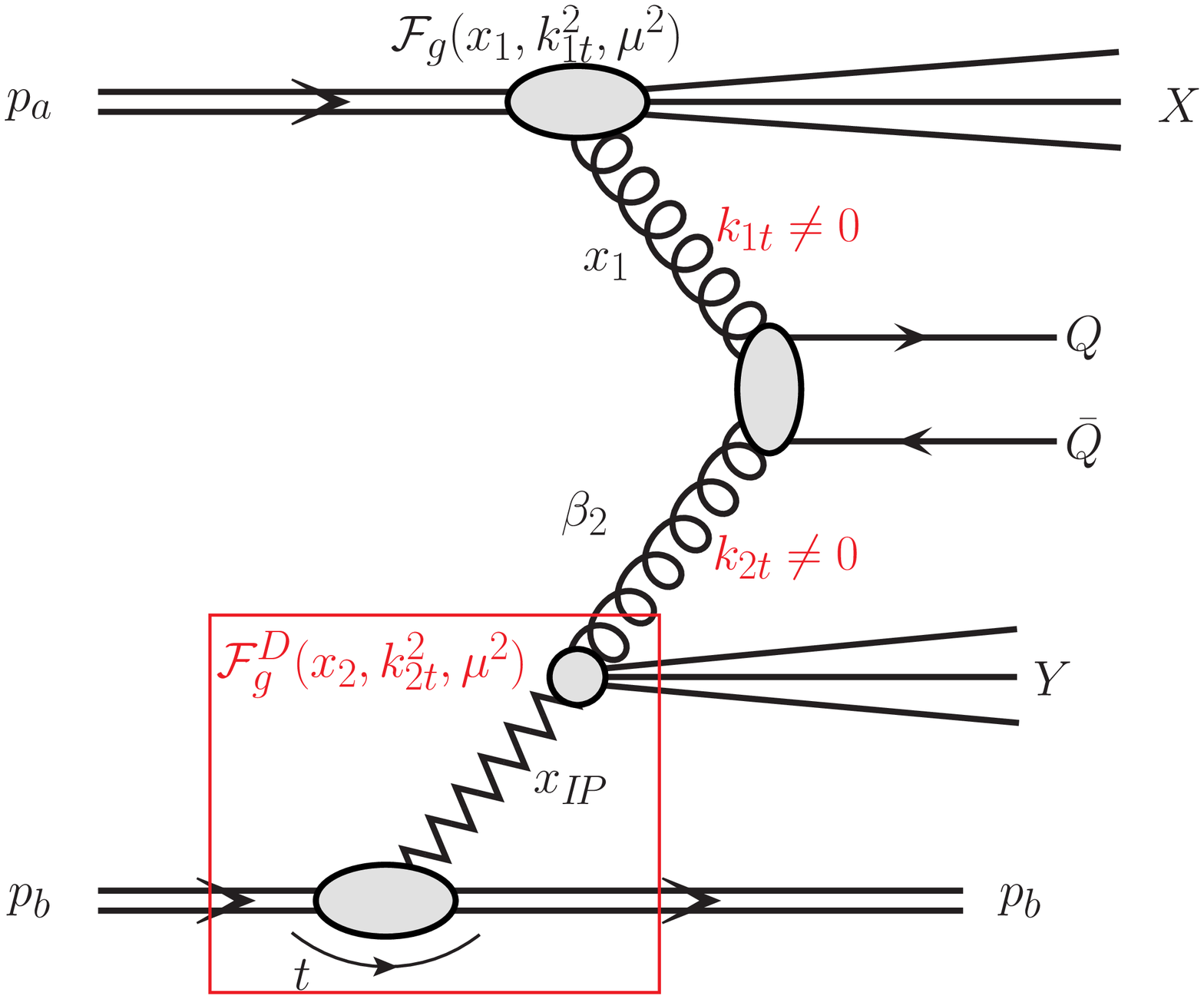}}
\end{minipage}
\caption{
\small A diagrammatic representation for single-diffractive production of heavy quark pairs within the $k_{t}$-factorization approach.
}
 \label{fig:mechanism}
\end{figure}

A sketch of the theoretical formalism is shown in Fig.~\ref{fig:mechanism}. Here, extension of the standard resolved pomeron model based on the LO collinear approach by adopting a framework of the $k_{t}$-factorization is proposed as an effective way to include higher-order corrections. According to this model the cross section for a single-diffractive production of charm quark-antiquark pair, for both considered diagrams (left and right diagram of Fig.~\ref{fig:mechanism}), can be written as:
\begin{eqnarray}
d \sigma^{SD(a)}({p_{a} p_{b} \to p_{a} c \bar c \; X Y}) &=&
\int d x_1 \frac{d^2 k_{1t}}{\pi} d x_2 \frac{d^2 k_{2t}}{\pi} \; d {\hat \sigma}({g^{*}g^{*} \to c \bar c }) \nonumber \\
&& \times \; {\cal F}_{g}^{D}(x_1,k_{1t}^2,\mu^2) \cdot {\cal F}_{g}(x_2,k_{2t}^2,\mu^2) ,
\label{SDA_formula}
\end{eqnarray}
\begin{eqnarray}
d \sigma^{SD(b)}({p_{a} p_{b} \to c \bar c p_{b} \; X Y}) &=&
\int d x_1 \frac{d^2 k_{1t}}{\pi} d x_2 \frac{d^2 k_{2t}}{\pi} \; d {\hat \sigma}({g^{*}g^{*} \to c \bar c }) \nonumber \\
&& \times \; {\cal F}_{g}(x_1,k_{1t}^2,\mu^2) \cdot {\cal F}_{g}^{D}(x_2,k_{2t}^2,\mu^2),
\label{SDB_formula}
\end{eqnarray}
where ${\cal F}_{g}(x,k_{t}^2,\mu^2)$ are the unintegrated ($k_{t}$-dependent) gluon distributions (UGDFs) in the proton and ${\cal F}_{g}^{D}(x,k_{t}^2,\mu^2)$ are their diffractive counterparts -- diffractive UGDFs (dUGDFs).

Details of our new calculations can be found in Ref.~\cite{Luszczak:2016csq}.

\section{Results}

\begin{figure}[!htbp]
\begin{minipage}{0.47\textwidth}
 \centerline{\includegraphics[width=1.0\textwidth]{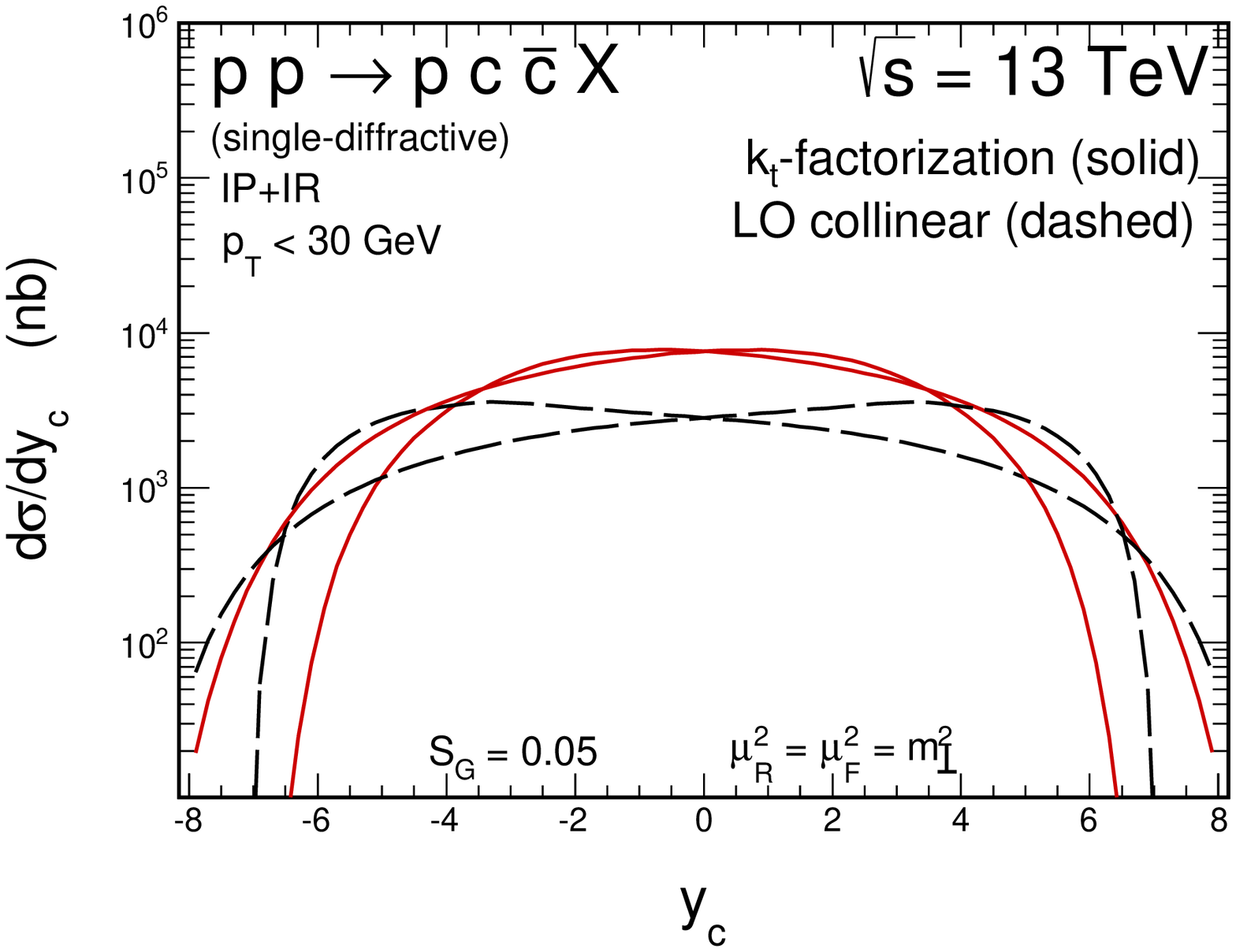}}
\end{minipage}
\hspace{0.5cm}
\begin{minipage}{0.47\textwidth}
 \centerline{\includegraphics[width=1.0\textwidth]{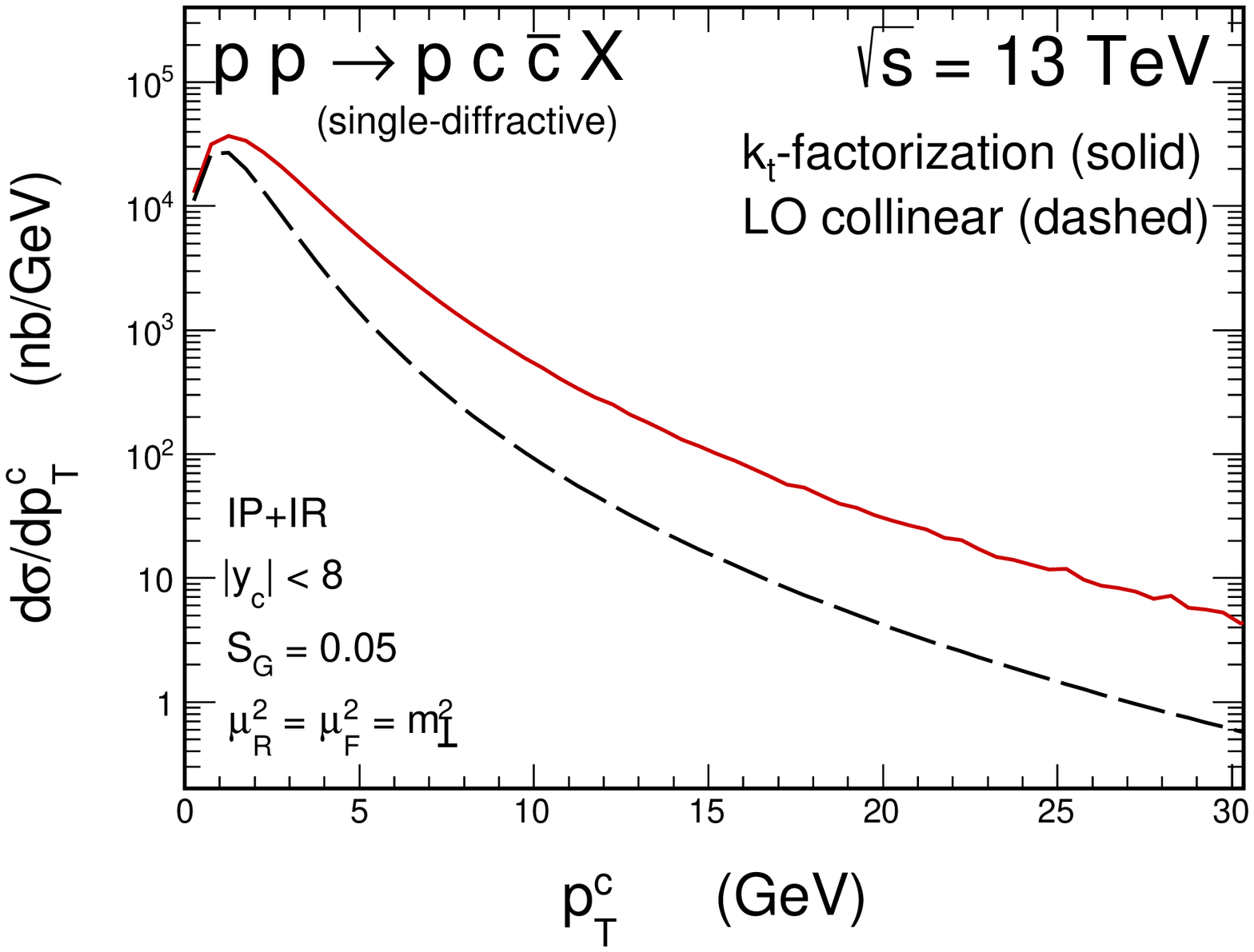}}
\end{minipage}
   \caption{
\small Rapidity (left panel) and transverse momentum (right panel) distributions of $c$ quarks (antiquarks) for a single-diffractive production at $\sqrt{s} = 13$ TeV. Components of the $g(I\!P)\operatorname{-}g(p)$, $g(p)\operatorname{-}g(I\!P)$, $g(I\!R)\operatorname{-}g(p)$, $g(p)\operatorname{-}g(I\!R)$ mechanisms are shown. 
}
 \label{fig:ypt_kTcoll}
\end{figure}

In Fig.~\ref{fig:ypt_kTcoll} we show rapidity (left panel) and transverse momentum (right panel) distribution of $c$ quarks (antiquarks) for single diffractive production at $\sqrt{s} = 13$ TeV.
Distributions calculated within the LO collinear factorization (black long-dashed lines) and for the $k_{t}$-factorization approach (red solid lines) are shown separately. We see significant differences between the both approaches, which are consistent with the conclusions from similar studies of standard non-diffractive charm production (see \textit{e.g}. Ref.~\cite{Maciula:2013wg}). Here we confirm that the higher-order corrections are very important also for the diffractive production of charm quarks.

The correlation observables can not be calculated within the LO collinear factorization but can be
directly obtained in the $k_{t}$-factorization approach.
\begin{figure}[!htbp]
\begin{minipage}{0.47\textwidth}
 \centerline{\includegraphics[width=1.0\textwidth]{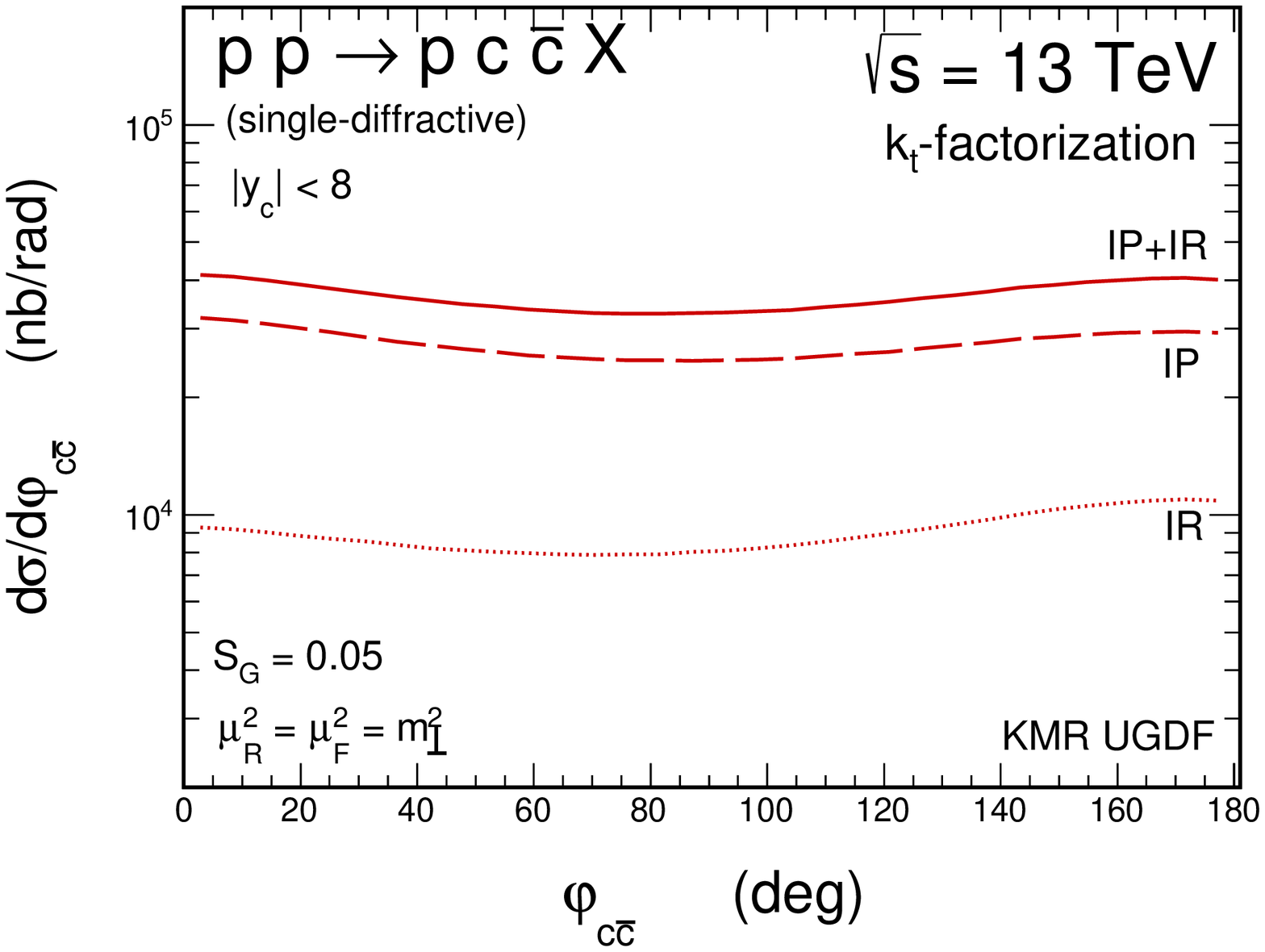}}
\end{minipage}
\hspace{0.5cm}
\begin{minipage}{0.47\textwidth}
 \centerline{\includegraphics[width=1.0\textwidth]{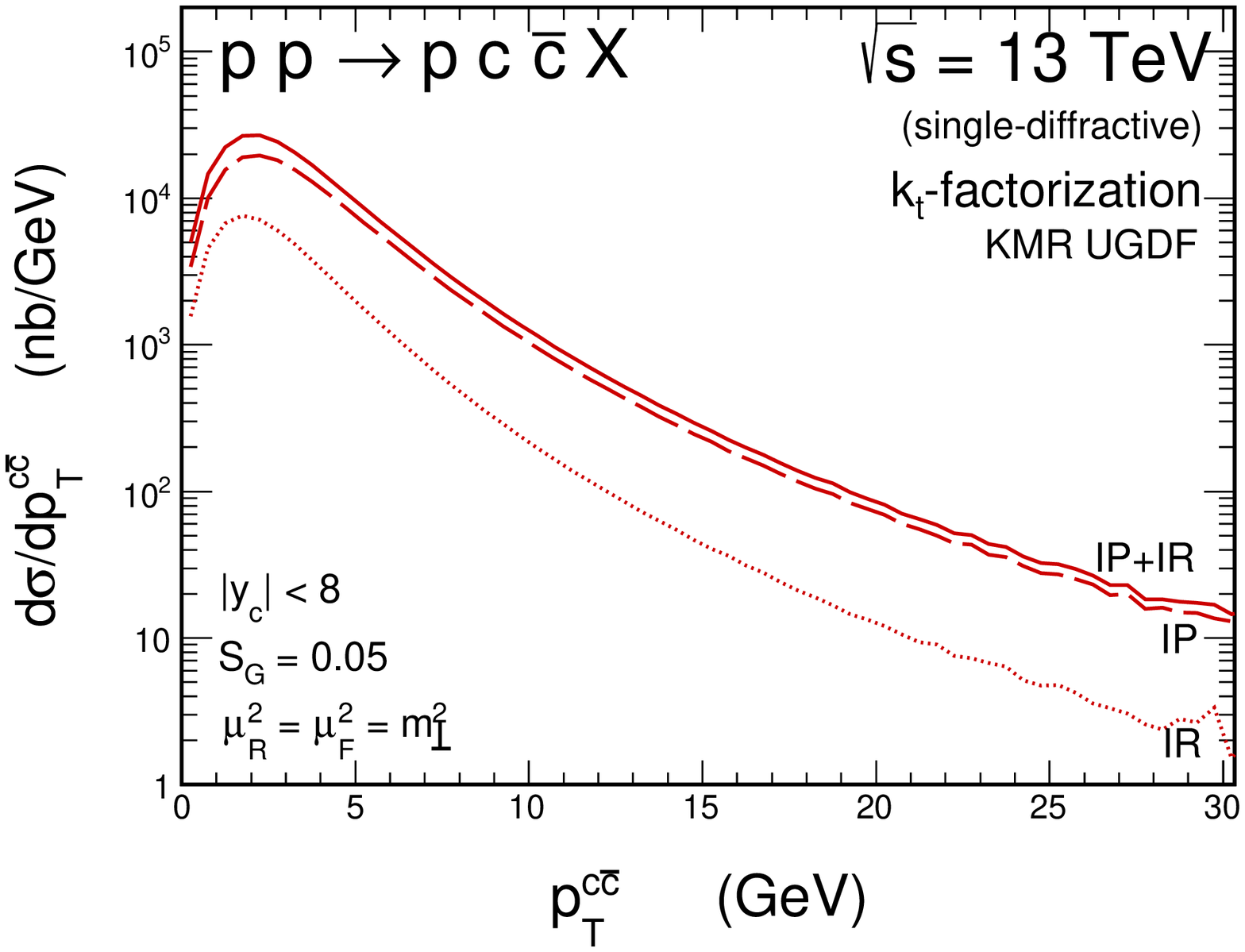}}
\end{minipage}
   \caption{
\small The distribution in $\phi_{c \bar c}$ (left panel) and distribution in $p_{T}^{c\bar c}$ 
(right panel) for $k_{t}$-factorization approach at $\sqrt{s}$ = 13 TeV.
}
 \label{fig:phid_ptsum_kT_PR}
\end{figure}
The distribution of azimuthal angle $\varphi_{c \bar c}$ between $c$ quarks and $\bar c$ antiquarks is shown in the left panel of Fig.~\ref{fig:phid_ptsum_kT_PR}. The $c \bar c$ pair transverse momentum distribution $p^{c \bar c}_{T} = |\vec{p^{c}_{t}} + \vec{p^{\overline{c}}_{t}}|$ is shown on the right panel. Results of the full phase-space calculations illustrate that the quarks and antiquarks in the $c \bar c$ pair are almost uncorrelated in the azimuthal angle between them and are often produced in the configuration with quite large pair transverse momenta. 
\begin{figure}[!htbp]
\begin{minipage}{0.4\textwidth}
 \centerline{\includegraphics[width=1.0\textwidth]{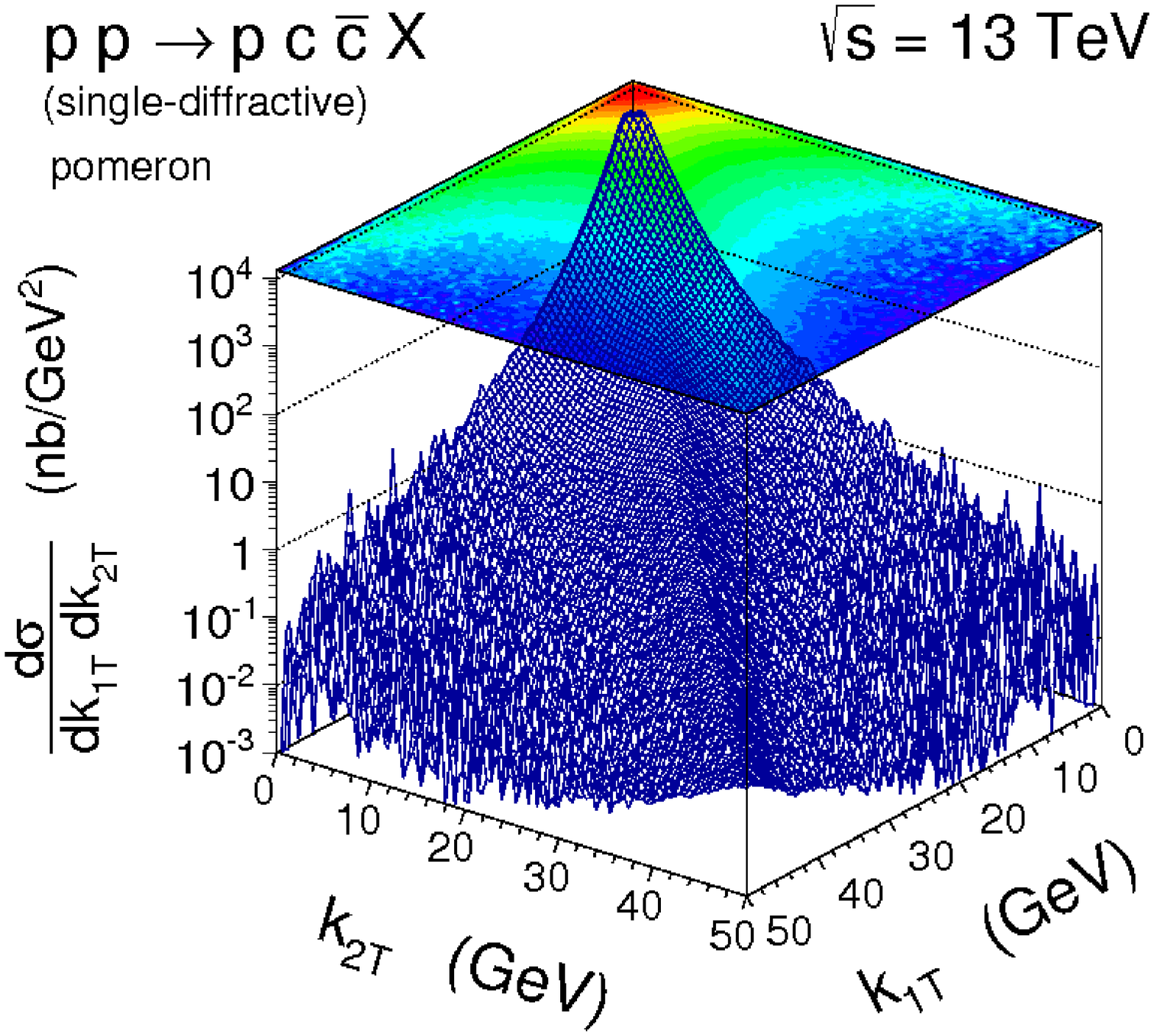}}
\end{minipage}
\hspace{0.5cm}
\begin{minipage}{0.4\textwidth}
 \centerline{\includegraphics[width=1.0\textwidth]{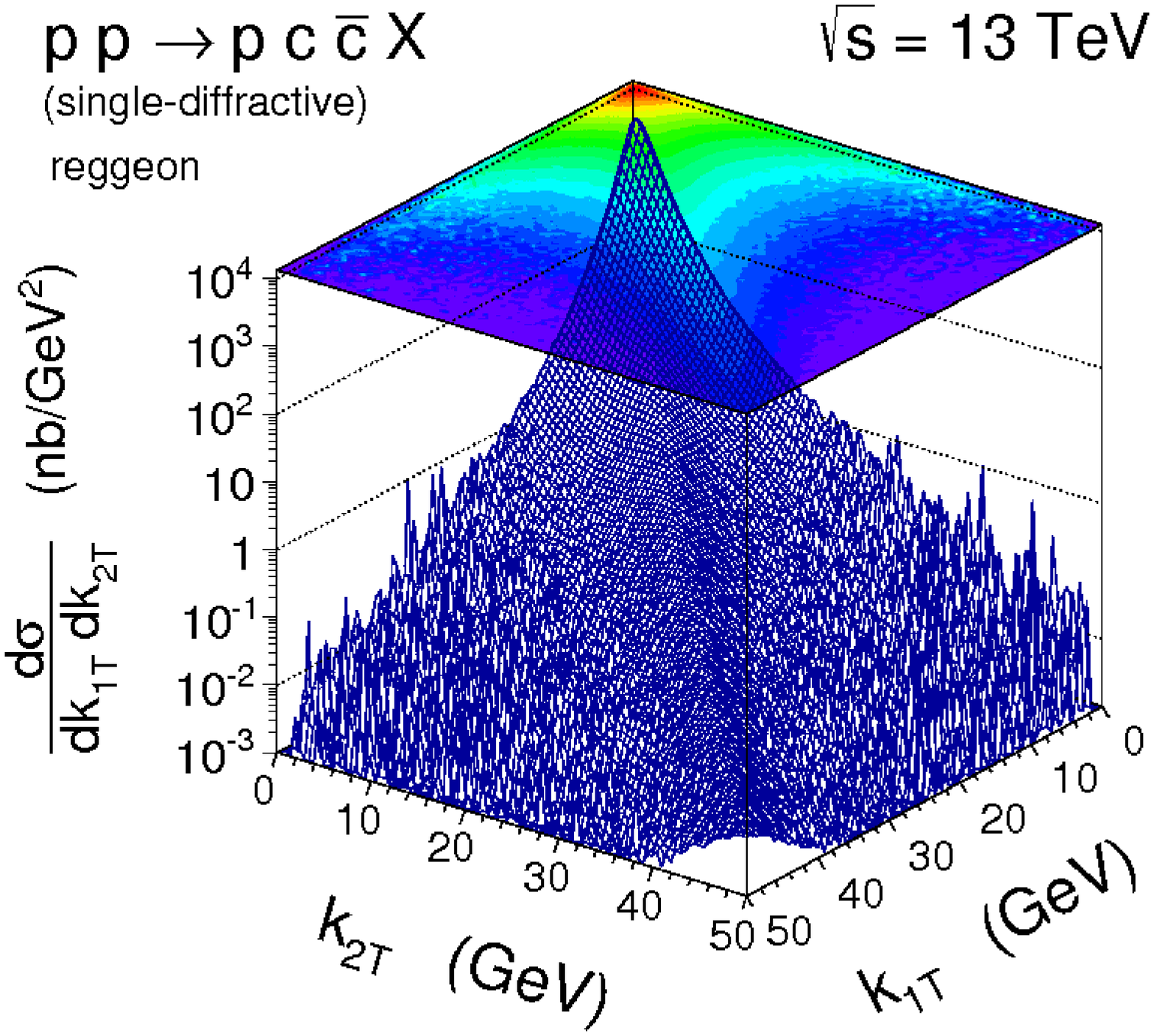}}
\end{minipage}
   \caption{
\small Double differential cross sections as a function of initial gluons transverse momenta $k_{1T}$ and $k_{2T}$ for single-diffractive production of charm at $\sqrt{s}=13$ TeV. The left and right panels correspond to the pomeron and reggeon exchange mechanisms, respectively. 
}
 \label{fig:q1tq2t_kT_PR}
\end{figure}
\begin{figure}[!htbp]
\begin{minipage}{0.4\textwidth}
 \centerline{\includegraphics[width=1.0\textwidth]{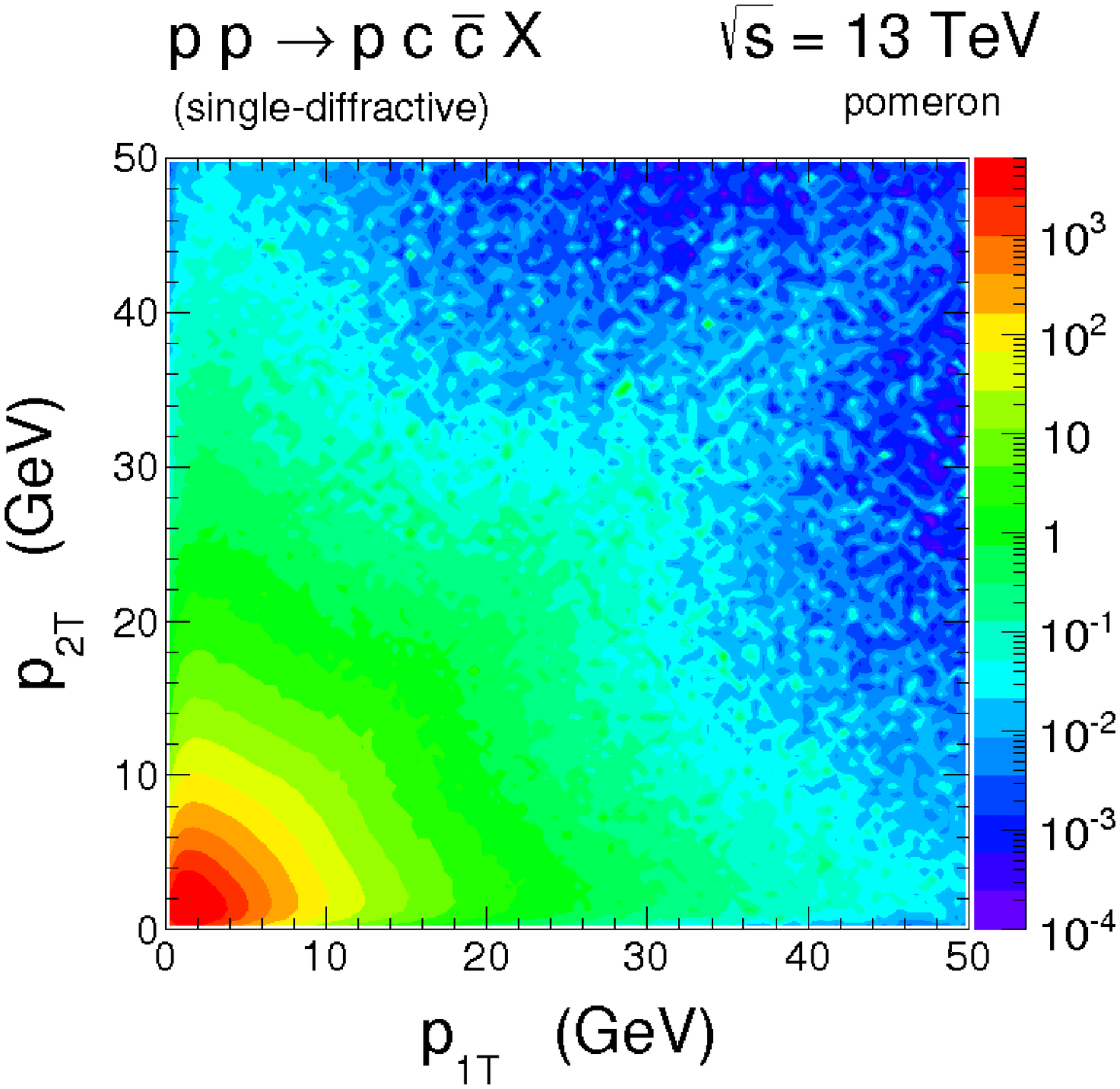}}
\end{minipage}
\hspace{0.5cm}
\begin{minipage}{0.4\textwidth}
 \centerline{\includegraphics[width=1.0\textwidth]{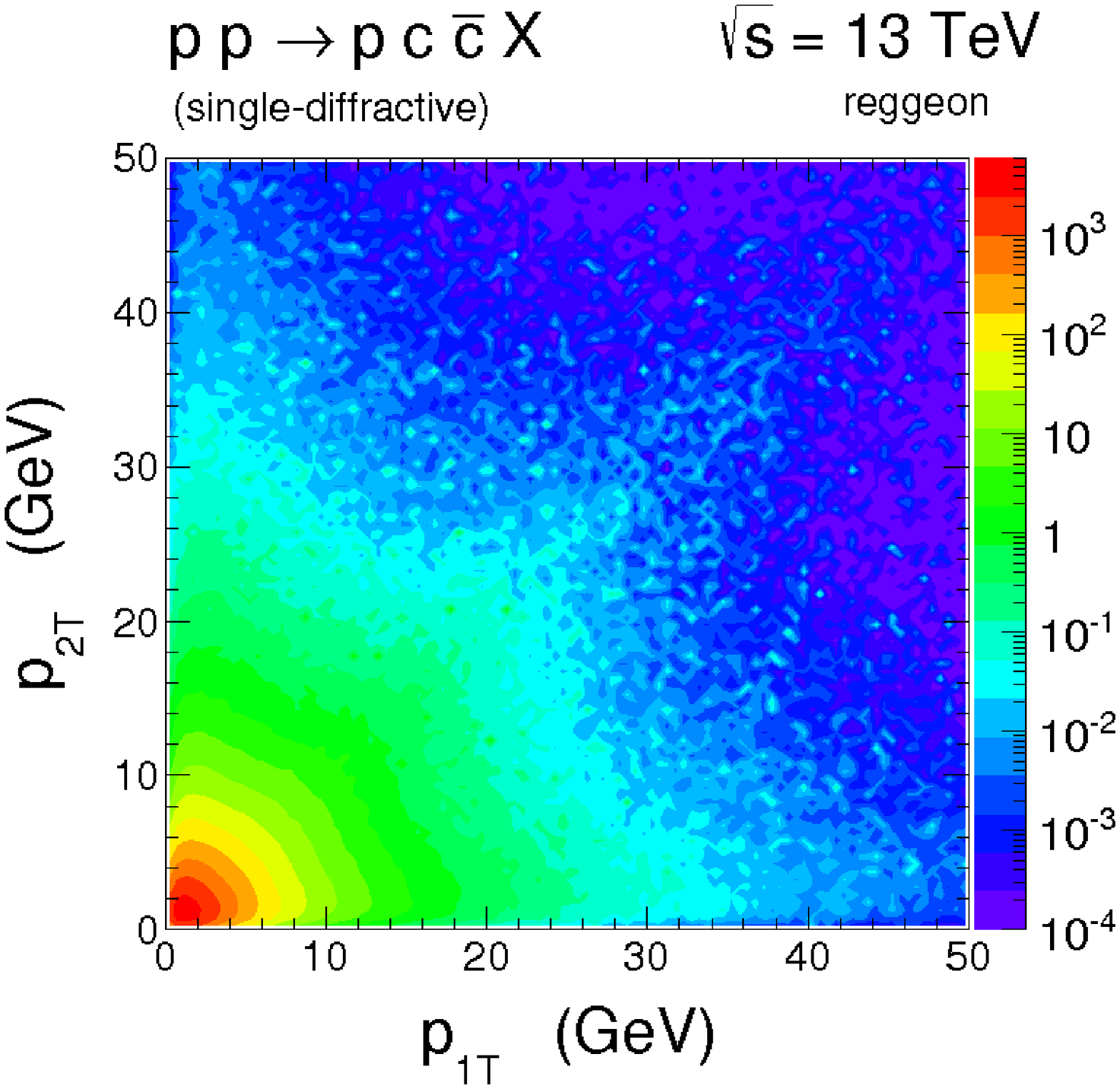}}
\end{minipage}
   \caption{
\small  Double differential cross sections as a function of transverse momenta of outgoing $c$ quark $p_{1T}$ and outgoing $\bar c$ antiquark $p_{2T}$ for single-diffractive production of charm at $\sqrt{s}=13$ TeV. The left and right panels correspond to the pomeron and reggeon exchange mechanisms, respectively. 
}
 \label{fig:p1tp2t_kT_PR}
\end{figure}
Figures~\ref{fig:q1tq2t_kT_PR} and \ref{fig:p1tp2t_kT_PR} show the double differential cross sections as a functions of transverse momenta of incoming gluons ($k_{1T}$ and $k_{2T}$) and transverse momenta of outgoing $c$ and $\bar c$ quarks ($p_{1T}$ and $p_{2T}$), respectively. We observe quite large incident gluon transverse momenta. The major part of the cross section is concentrated in the region of small $k_{t}$'s of both gluons but long tails are present. Transverse momenta of the outgoing particles are not balanced as they were in the case of the LO collinear approximation.
\section{Conclusions}
Charm production is a good example where the higher-order effects are very important. For the inclusive charm production we have shown that these effects can be effectively included in the $k_t$-factorization approach \cite{Maciula:2013wg}. 
In our approach we decided to use the so-called KMR method to calculate unintegrated diffractive gluon distribution. As usually in the KMR approach, we have calculated diffractive gluon UGDFs based on collinear distribution, which in the present case is diffractive collinear gluon distribution. In our calculations we have used the H1 Collaboration parametrization fitted to the HERA data on diffractive structure function and di-jet 
production.
Having obtained unintegrated diffractive gluon distributions we have
performed calculations of several single-particle and correlation
distributions. In some cases the results have been compared with the
results obtained in the leading-order collinear approximation. In
general, the $k_t$-factorization approach leads to larger cross
section. However, the $K$-factor is strongly dependent on phase space
point. Some correlation observables, like azimuthal angle correlation
between $c$ and $\bar c$, and $c \bar c$ pair transverse momentum were
obtained in \cite{Luszczak:2016csq} for the first time.


\end{document}